# n-Dimensional Gravity:
# Little Black Holes, Dark Matter, and Ball Lightning

Mario Rabinowitz
Armor Research; lrainbow@stanford.edu
715 Lakemead Way, Redwood City, CA 94062-3922

**Abstract**

The gravitational field, and radiation from quantized gravitational atoms and little black holes (LBH) are analyzed in n-space, i.e. in all dimensions from 0 to ∞ to develop insights into possible additional compacted dimensions as predicted by hierarchy and string theory. It is shown that the entropy of LBH is significantly greater in higher dimensional space with potential implications to the initial entropy of the universe. A case is made that LBH are the dark matter of the universe, and can manifest themselves as the core energy source of ball lightning (BL). The LBH incidence rate on earth is related to BL occurrence and has the potential of aiding in the determination of the distribution of LBH and hence dark matter in the universe. Possibilities are explored as to why Hawking radiation has been undetected in over 25 years. An alternate LBH tunneling radiation model is described.

## 1. INTRODUCTION

Much effort is underway both experimentally and theoretically to probe the physical implications of potentially higher compacted dimensions at accessible energies. This includes testing for a deviation from the Newtonian $1/r^2$ gravitational force at sub-millimeter distances due to additional dimensions as predicted by a combination of hierarchy and string theory. One object of this paper is to analyze the properties of little black holes (LBH) in all dimensions as they may be the only highly massive bodies capable of existing inside these compacted dimensions. Quantized gravitational orbits with LBH that can exist inside compacted dimensions and radiate gravitons (gravitational radiation) are examined.

This is not to be viewed as an exercise in metaphysics, since compacted dimensions "could have physical meaning" as Kaluza-Klein theory shows (Finkelstein, 1997). The word "little" as used here with respect to a black hole refers to its radius rather than its mass since a little black hole (LBH) the size of a neutron (~ $10^{-15}$ m) weighs as much as a mountain ($10^{12}$ kg    $10^9$ ton) and a $10^{-3}$ kg LBH has a radius of $10^{-30}$ m.    Possibilities are explored as to why Hawking radiation has been undetected. One approach assumes the correctness of the model, but analyses the effects of additional compact dimensions on the attenuation of this radiation. Another considers the prospect that the analysis supporting the Hawking model is flawed. An alternate radiation tunneling model is described.

A case is made that LBH are excellent candidates for the missing mass of the universe. A noteworthy finding, possibly related to the initial entropy of the universe, is that the entropy of LBH is significantly greater in higher dimensional space. In any case, LBH may help to solve the problem of why the early universe appears to have too little entropy.

## 2. GRAVITATIONAL FIELD AND POTENTIAL ENERGY IN n-SPACE
### 2.1 n-Dimensional Space

Gauss's law in three-dimensional Euclidean space implies Newton's inverse square gravitational force law, $F = \dfrac{GMm}{r^2}$, and in general the power of r is (n-1), where n = 3, 4, 5, ... is the number of spatial dimensions in the space-time manifold of (n+1) dimensions. Gauss's law for the gravitational field in n-space is

$$\oint \vec{F}_n \bullet d\vec{A}_{n-1} = \oint 4\pi G_n \rho_n m dV_n = 4\pi G_n Mm, \qquad (1)$$

where M is the mass of density $\rho_n$ enclosed inside the (n-1)-dimensional area $A_{n-1}$, and m is a test mass in the field produced by M. The universal gravitational constant $G_n$ may be expected to increase (in a way that is model dependent) from its 3-space value as the number of dimensions increases if the higher dimensions are compacted as in string theory. In general for spherical symmetry where $F_n$ is normal to $A_{n-1}$ and constant over its surface, eq. (1) implies

$$F_n = \frac{4\pi G_n M m}{A_{n-1}} \qquad (2)$$

Let us consider a spherical surface

$$A_{sn-1} = S_{n-1} r^{n-1} \qquad (3)$$

of an n-dimensional hypersphere (n-sphere, or more precisely n-ball) ) of volume, $V_{sn} = C_n r^n$, where $S_{n-1}$ and $C_n$ are numerical coefficients which are next calculated.

$$\frac{dV_{sn}}{dr} = nC_n r^{n-1} = A_{sn-1} \Rightarrow nC_n = S_{n-1} \qquad (4)$$

Given that $\int_{-\infty}^{\infty} e^{-x^2} dx = 2\int_0^{\infty} e^{-x^2} dx = \pi^{1/2}$, this implies that

$$\left[\pi^{1/2}\right]^n = \left[\int_{-\infty}^{\infty} e^{-x^2} dx\right]^n = \iiint \ldots e^{-\sum_i^n x_i^2} dx_1 dx_2 \ldots dx_n. \qquad (5)$$

Converting from Cartesian to polar coordinates in n-space

$$r^2 = \sum_{i=1}^n x_i^2, \text{ and } dx_1 dx_2 \ldots dx_n = r^{n-1} d\varphi_1 d\varphi_2 \ldots d\varphi_{n-1} dr. \qquad (6)$$

Substituting eq. (6) into (5) yields

$$\pi^{n/2} = \int_0^{\infty} \iiint \ldots e^{-r^2} r^{n-1} d\varphi_1 d\varphi_2 \ldots d\varphi_{n-1} dr. \qquad (7)$$

Now

$$\iiint \ldots d\varphi_1 d\varphi_2 \ldots d\varphi_{n-1} = S_{n-1}, \qquad (8)$$

since $V_{sn} = \int_0^R S_{n-1} r^{n-1} dr = \int_0^R \iiint \ldots r^{n-1} d\varphi_1 d\varphi_2 \ldots d\varphi_{n-1} dr. \qquad (9)$

Substituting eqs. (8) and (9) into (7) yields

$$\pi^{n/2} = \int_0^{\infty} S_{n-1} e^{-r^2} r^{n-1} dr. \qquad (10)$$

Let $t = r^2$, $\Rightarrow dr = \tfrac{1}{2} t^{-1/2} dt$, so eq. (10) becomes

$$\pi^{n/2} = \int_0^{\infty} S_{n-1} e^{-t} t^{(n-1)/2} \left(\tfrac{1}{2} t^{-1/2} dt\right)$$

$$= \frac{S_{n-1}}{2} \int_0^{\infty} t^{\left(\frac{n}{2}-1\right)} e^{-t} dt = \frac{S_{n-1}}{2} \Gamma\left(\frac{n}{2}\right), \qquad (11)$$

where the Gamma function $\Gamma(n) \equiv \int_0^\infty t^{n-1} e^{-t} dt$ for all n (integer and non-integer). When n is an integer, $\Gamma(n) = (n-1)!$ Thus eq. (11) implies $S_{n-1} = 2\pi^{n/2} / \Gamma\left(\dfrac{n}{2}\right)$ and by eq. (2) the (n-1)-area of an n-sphere is

$$A_{sn-1} = \frac{2\pi^{n/2}}{\Gamma\left(\dfrac{n}{2}\right)} r^{n-1}. \tag{12}$$

By eq. (4),
$$C_n = \frac{S_{n-1}}{n} = 2\pi^{n/2} \Big/ n\Gamma\left(\frac{n}{2}\right) = \pi^{n/2} \Big/ \frac{n}{2}\Gamma\left(\frac{n}{2}\right) = \pi^{n/2} \Big/ \Gamma\left(\frac{n}{2}+1\right).$$ Substituting this into eq. (1) gives the n-volume of an n-sphere:

$$V_{sn} = \frac{\pi^{n/2}}{\Gamma\left(\dfrac{n}{2}+1\right)} r^n. \tag{13}$$

Subsequent to deriving the above results, I found a proof by Coxeter (1948) that has both similarities and differences with my derivation. Coxeter does not claim originality, nor does he cite a reference.

Equations (12) and (2) give the gravitational force in such an n-space,

$$F_n = \frac{-4\pi G_n Mm}{A_{n-1}} = \frac{-2\pi G_n Mm \Gamma\left(\dfrac{n}{2}\right)}{\pi^{n/2} r^{n-1}}, \tag{14}$$

where the gravitational constant $G_n$ is model dependent for n > 3. Thus the gravitational potential energy in n dimensions is

$$\Phi_n = -\int \vec{F}_n \bullet d\vec{r} = \frac{-2\pi G_n Mm \Gamma\left(\dfrac{n}{2}\right)}{(n-2)\pi^{n/2} r^{n-2}}. \tag{15}$$

Measurements on the cosmic background microwave radiation indicate that on a large scale our 3-space universe is Euclidean so the above calculations appear relevant as an extrapolation to a potentially physical Euclidean n-space. In Section 5, the domain of applicability of Newtonian gravity is examined.

## 2.2 Infinite Dimensional Space

Interestingly, from eqs. (12) and (13) we find that for integer n, the n-1 area of an n-sphere peaks in 7-dimensional space and the n-volume of an n-sphere (n-ball) relative to an n-cube of side r, peaks in 5-dimensional space. Thereafter both decrease monotonically to zero. This can be discerned by examining the limit $n \to \infty$ for the n-volume of an n-sphere of fixed radius r, in n-space as given by eq. (13). Though a more complicated and more accurate approximation can be used for the Gamma function, we use the standard Stirling approximation for simplicity:

$$V_{sn} = \frac{\pi^{n/2}}{\Gamma\left(\frac{n}{2}+1\right)} r^n \approx \frac{\pi^{n/2} r^n}{\frac{n}{2} e^{\frac{-n}{2}} \left(\frac{n}{2}\right)^{\frac{n}{2}} \sqrt{2\pi\left(\frac{n}{2}\right)}} \underset{n \to \infty}{\to} 0 \quad . \tag{16}$$

This result may seem particularly surprising in view of the fact that the volume of an n-space hypercube (n-cube) of fixed side $r > 1$,

$$V_{cn} = r^n \underset{n \to \infty}{\to} \infty. \tag{17}$$

Taking $r < 1$, so that $r^n \underset{n \to \infty}{\to} 0$ would not really resolve this result. To avoid scale problems we can make our result scale independent, and only a little less surprising. Thus:

$$\frac{V_{sn}}{V_{cn}} = \frac{\pi^{n/2}}{\Gamma\left(\frac{n}{2}+1\right)} \approx \frac{\pi^{n/2}}{\frac{n}{2} e^{\frac{-n}{2}} \left(\frac{n}{2}\right)^{\frac{n}{2}} \sqrt{2\pi\left(\frac{n}{2}\right)}} \underset{n \to \infty}{\to} 0. \tag{18}$$

This interesting result occurs because as the dimensionality increases, the main diagonal of the n-cube exceeds the diameter of the n-sphere, which remains fixed at 2r. Holding the n-cube side length fixed at r, the volume of the n-cube increases faster than the volume of the n-sphere as n increases. In 4-space, a 4-sphere circumscribes a 4-cube since the main diagonal of the 4-cube is $\sqrt{4}\, r$ = 2r = diameter of the 4-sphere. In n-space, the main diagonal of the n-cube is $\sqrt{n}\, r$.

Another way to look at this is to circumscribe an n-cube of side 2r around an n-sphere of radius r. The number of corners is $2^n$, and as n gets arbitrarily large, the n-cube becomes "all corners." The distance from the center of the n-cube to one of its corners is $\sqrt{n}\,r$. Whereas the sphere extends only the fractional distance $r/\sqrt{n}\,r = 1/\sqrt{n}$ toward the n-cube's corners. For large n, the ratio of the n-sphere volume to the n-cube volume is a quickly diminishing fraction $< r^n/(2r)^n = 1/2^n \xrightarrow[n\to\infty]{} 0$.

## 3. n-DIMENSIONAL QUANTIZED GRAVITATIONAL ORBITS

Let us next consider quantized non-relativistic gravitational orbits in n-space. These would be the analog of electrostatic atomic orbitals. Ordinary matter does not have a high enough density to make such orbits feasible, but LBH do. For example in 3-space, a $10^{-3}$ kg (1 gram) LBH with a radius of $10^{-30}$ m has a density $> 10^{86}$ kg/m$^3$, whereas nucleon densities are only $\sim 10^{18}$ kg/m$^3$.

One might challenge the use of semi-classical physics at such a small scale and high energies. However, as measured at large distances, the gravitational red shift substantially reduces the impact of the high energies near LBH. Furthermore, Argyres et al (1998) argue that "...one can describe black holes by semi-classical physics down to much smaller masses of order the fundamental Planck scale... ," where the Planck mass of $10^{-8}$ kg has a LBH radius of $10^{-35}$ m.

Bohr-Sommerfeld semi-classical theory is not only less accurate, it is less consistent than quantum theory (Finkelstein,1997). Furthermore, orbital motion is non-reentrant when the force law differs from $1/r^2$ as indicated by the Runge vector (or Runge-Lenz vector, quantum mechanically). Nevertheless as an approximation to simplify the analysis, we will use the Bohr-Sommerfeld condition $\oint p_\phi d\phi = j\hbar$, $j = 1, 2, 3, ...$ which works well for hydrogen-like circular orbits. Its use for gravitational orbits has interesting consequences (Rabinowitz, 1990 a).

For $M \gg m$, we have from eq. (14)

$$F_n = \frac{2\pi G_n Mm \Gamma\left(\dfrac{n}{2}\right)}{\pi^{n/2} r^{n-1}} = \frac{mv_n^2}{r_n} \qquad (19)$$

Equation (19) implies that the n-space orbital velocity is

$$v_n = \left[\frac{2\pi G_n M \Gamma\left(\frac{n}{2}\right)}{\pi^{n/2} r^{n-2}}\right]^{1/2} . \qquad 20)$$

From the Bohr-Sommerfeld condition, $mv_n r_n = j\hbar$, we find for the orbital radius of m around M

$$r_n = \left[\frac{j\hbar \pi^{\frac{n-2}{4}}}{m[2G_n M \Gamma(n/2)]^{1/2}}\right]^{\frac{2}{4-n}} . \qquad (21)$$

In 3-space eq. (21) yields $r_3 = \frac{j^2 \hbar^2}{GMm^2}$.

Substituting eq. (21) into eq. (20), the orbital velocity is

$$v_n = \left\{\left[\frac{2\pi G_n M \Gamma\left(\frac{n}{2}\right)}{\pi^{n/2}}\right]\left[\frac{m^{\frac{2}{4-n}}\left[2G_n M \Gamma\left(\frac{n}{2}\right)\right]^{\frac{1}{4-n}}}{(j\hbar)^{\frac{2}{4-n}} \pi^{\frac{n-2}{2(4-n)}}}\right]\right\}^{1/2} . \qquad (22)$$

In 3-dimensions eq. (22) gives $v_3 = \frac{GMm}{j\hbar}$.

Using equations (14) and (21), the acceleration of the orbiting mass m is

$$a_n = \frac{F_n}{m} = \frac{-2\pi G_n M \Gamma\left(\frac{n}{2}\right)}{\pi^{n/2}} \left[\frac{m[2G_n M \Gamma(n/2)]^{1/2}}{j\hbar \pi^{\frac{n-2}{4}}}\right]^{\frac{2n-2}{4-n}} . \qquad (23)$$

In 3-dimensional space, eq. (23) yields $a_3 = \frac{-G^3 M^3 m^4}{(j\hbar)^4}$. This is a violation of the equivalence principle, because the acceleration is not independent of m, is not an artifact of the Bohr-Sommerfeld condition. Quantum mechanical interference effects in general and quantum-gravitational interference effects in particular depend on the phase which depends on the mass.

In n-space, the total energy of the gravitationally bound atom of nucleus mass M and orbiting mass m is obtained from equations (22) and (15):

$$E_n = \tfrac{1}{2}mv_n^2 + \Phi_n = \frac{m(n-4)}{n-2}\left[\frac{G_n M \Gamma\left(\frac{n}{2}\right)}{\pi^{\frac{n-2}{2}}}\right]\left[\frac{m^{\frac{2(n-2)}{4-n}}\left[2G_n M \Gamma\left(\frac{n}{2}\right)\right]^{\frac{n-2}{4-n}}}{(j\hbar)^{2(n-2)/(4-n)}\pi^{(n-2)^2/2(4-n)}}\right]. \quad (24)$$

In 3-space, eq. (24) reduces to $E_3 = -\dfrac{m^3}{2}\left(\dfrac{G_3 M}{j\hbar}\right)^2$. The binding energy between M and m of such an atom is given by j = 1, and is generally small, unless M and m are large compared to elementary particle masses. The binding energy can be much greater in higher dimensional space.

The frequency of radiated gravitons is
$$\nu_n = \frac{\Delta E}{2\pi\hbar}$$

$$\approx \frac{m(n-4)}{2\pi\hbar(n-2)}\left[\frac{G_n M \Gamma\left(\frac{n}{2}\right)}{\pi^{\frac{n-2}{2}}}\right]\left[\frac{m^{\frac{2(n-2)}{4-n}}\left[2G_n M \Gamma\left(\frac{n}{2}\right)\right]^{\frac{n-2}{4-n}}}{(\hbar)^{2(n-2)/(4-n)}\pi^{(n-2)^2/2(4-n)}}\right]\left[\frac{1}{(j+2)^{2(n-2)/(4-n)}} - \frac{1}{(j)^{2(n-2)/(4-n)}}\right]. \quad (25)$$

The principal quantum number j changes to j+2 to conserve angular momentum for the system of atom and emitted graviton of spin 2. For a quadrupole transition of a gravitationally bound atom in 3-space, the emitted graviton frequency for de-excitation between nearest allowed energy states is

$$\nu_3 = \frac{m^3(G_3 M)^2}{4\pi\hbar^3}\left[\frac{1}{j^2} - \frac{1}{(j+2)^2}\right]. \quad (26)$$

## 4. GRAVITATIONAL RADIATION

Preparation is being made by laboratories to measure gravitational radiation from distant sources. Such experiments are being conducted by teams around the world: U.S. Laser Interferometer Gravitational-Wave Observatory (LIGO); VIRGO (France/Italy); GEO-600 (Britain/ Germany); TAMA (Japan); and ACIAGA (Australia).

The detectors are laser interferometers with a beam splitter and mirrors suspended on wires. The predicted gravitational wave displaces the mirrors and shifts the relative optical phase in two perpendicular paths. This causes a shift in the interference pattern at the beam splitter. It is expected that by 2010, the devices will be sensitive enough to detect gravitational waves up to $10^2$ Megaparsecs ( 3.26 x $10^8$ lightyear = 3.1 x $10^{24}$ m). A major challenge has arisen because the detector noise does not satisfy the usual assumptions of being stationary and Gaussian (Allen et al, 1999).

As shown in section 3, gravitational radiation is possible from gravitationally bound atoms. We may inquire whether a signal from such potentially nearby sources can compete or interfere with distant sources such as neutron stars, binary pulsars, and coalescing black holes. Signals from such sources are expected to have frequencies in the range from 10 Hz to $10^4$ Hz (Davies, 1992).

From eq. (26), we see that gravitational radiation from orbital de-excitation of a nucleon mass m ~ $10^{-27}$ kg orbiting a LBH of mass M ~ 10 kg, would emit a frequency ~ $10^3$ Hz. With a high enough concentration of excited orbits near the earth, the signals from LBH quantized gravitational orbits might compete with very distant signals. Such concentrations are unlikely for various reasons including the very low 3-dimensional binding energy of such orbits. Furthermore, a gravitational wave has details about its source, including its consistency with general relativity. Based upon these considerations, it appears unlikely that gravitational radiation from distant sources will be masked by potential nearby sources such as gravitational atoms.

## 5. SCHWARZSCHILD (HORIZON) RADIUS IN n-SPACE

Equations (14) and (15) should be generally valid for Euclidean n-space. In 3-dimensions the difference between the potential energy in Einstein's general relativity and Newtonian gravitation gets small for radial distances r > 10 $R_H$, where $R_H$ is the Schwarzchild (horizon) radius of a black hole. In 3-space $R_H$ = $2GM/c^2$, where M is the mass of a black hole and c is the speed of light.

This approximation should be valid for all scales since

$$V = \frac{G(M)\gamma m}{r} < \frac{G(R_H c^2/2G)\gamma m}{10 R_H} = \frac{\gamma m c^2}{20} \quad (27)$$

is scale independent, where $\gamma = (1 - v^2/c^2)^{-1/2}$. For velocities $v \ll c$, it is only necessary that V be smaller than 1/20 of the rest energy of the orbiting body. This approximation should be valid even for relativistic velocities when $\gamma$ is not negligible, since it cancels out on both sides of the inequality. As Finkelstein (1997) points out, when special relativity is used within the context of Newtonian gravity "the source of gravity ... [is] the inertial mass, not the rest mass."

So we may to a good approximation use the Newtonian potential for $r > 10 R_H$ for all sizes of black holes in 3-space. In higher dimensions, the results of the two theories should differ by only dimensionless numbers. Bear in mind that not only does Newton's law fail for $r < 10 R_H$, due to general relativity, but that both may fail, as does Coulomb's law, due to quantum effects at very small distances.

With eq. (15), we are now in a position to generalize the standard derivation of $R_H$ to n-space. By conservation of energy

$$\gamma m c^2 /2 + \frac{-4\pi G_n M(\gamma m) \Gamma\left(\frac{n}{2}\right)}{2(n-2)\pi^{n/2} R_{Hn}^{n-2}} = KE_\infty + PE_\infty = 0. \quad (28)$$

The kinetic energy term on the LHS is written here as it is in other conventional non-general relativistic derivations to avoid being a factor of 2 low for $R_H$ in 3-space. This simplified derivation is used to avoid undue complexity. It differs from general relativity in allowing emitted particles to be classically found at varying distances from a black hole. In general relativity, classically there are no emitted particles outside a black hole. Solving eq. (28) for the n-dimensional Schwarzchild radius

$$R_{Hn} = \left[\frac{4\pi G_n M \Gamma\left(\frac{n}{2}\right)}{(n-2)\pi^{n/2} c^2}\right]^{\frac{1}{n-2}}, \quad (29)$$

giving $R_{H3} = 2GM/c^2$ in 3-space. $R_{Hn} > R_{H3}$ for $n > 3$.

## 6. n-SPACE BLACK HOLE TEMPERATURE, RADIATED POWER

An intuitive insight may be gained by using the uncertainty principle to obtain the black hole temperature T, and Hawking radiated power $P_{SH}$. These heuristic results are illuminating and close to those achieved by more complex rigorous derivations despite neglect of important basic considerations such as gravitational redshift. They illustrate that $R_H$ is the key variable for n-space determination; that one need not invoke black-body radiation to obtain $P_{SH}$; and that $P_{SH}$ and T can be obtained independently of each other.

### 6.1 Black Hole Temperature from Uncertainty Principle

To obtain the BH temperature, let us use the uncertainty principle in a kind of scaled dimensional analysis. Though our main interest is in LBH because of their extremely high temperatures, where whatever is inside is ultra-relativistic so that v c, what we shall do is applicable to BH of all sizes.

$$T_n \approx \frac{\Delta E}{k} \sim \frac{(\hbar/2)/\Delta t}{k} \sim \frac{\hbar}{2k\left(\frac{2R_{Hn}}{v}\right)} \sim \frac{\hbar c}{4k\left[\frac{4\pi G_n M \Gamma\left(\frac{n}{2}\right)}{(n-2)\pi^{n/2} c^2}\right]^{\frac{1}{n-2}}}, \quad (30)$$

where k is the Boltzmann constant. In 3-space this reduces to

$$T_3 \sim \frac{\hbar c^3}{8kG}\left[\frac{1}{M}\right]. \quad (31)$$

This is just a factor of $\pi$ bigger than Hawking's T of (1974), and $2\pi$ bigger than his T of (1975).

### 6.2 Black Hole Radiated Power from Uncertainty Principle

Particles can borrow energy $\Delta E$ for a time $\Delta t$ to get over the top of the BH potential energy barrier. The more energy that needs to be borrowed, the sooner it has to be returned. In order for a particle to escape a BH, it must have this energy for a time at least long enough to cross the barrier.

$$P_{SHn} = \frac{\Delta E}{\Delta t} \sim \frac{(\hbar/2)/\Delta t}{\Delta t} = \frac{\hbar}{2(\Delta t)^2} = \frac{\hbar}{2\left\{\frac{2R_{Hn}}{c}\right\}^2}$$

$$= \frac{\hbar c^2}{8\left\{R_{Hn}^2\right\}} = \frac{\hbar c^2}{8\left\{\left[\frac{4\pi G_n M \Gamma\left(\frac{n}{2}\right)}{(n-2)\pi^{n/2} c^2}\right]^{\frac{2}{n-2}}\right\}} \quad , \tag{32}$$

In 3-space this reduces to

$$P_{SH3} \sim \frac{\hbar c^2}{2(2R_H)^2} = \frac{\hbar c^2}{8}\left[\frac{2GM}{c^2}\right]^{-2} = \frac{\hbar c^6}{32 G^2}\left[\frac{1}{M^2}\right]. \tag{33}$$

This is quite close to Hawking's radiated power (1974, 1975).

## 7. LITTLE BLACK HOLE TUNNELING RADIATION MODEL

Radiation may be emitted from black holes in a process differing from that of Hawking radiation, $P_{SH}$, which has been undetected for over 25 years. As derived in the Rabinowitz tunneling model (1999 a, b, c), beamed exhaust radiation $P_R$ tunnels out from a LBH due to the field of a second body, which lowers the LBH gravitational potential energy barrier and gives the barrier a finite width. Particles can escape by tunneling (as in field emission) or over the top of the lowered barrier (as in Schottky emission). The former is similar to electric field emission of electrons from a metal by the application of an external field.

Although $P_R$ is of a different physical origin than Hawking radiation, it is analytically of the same form since $P_R \propto \Gamma P_{SH}$, where $\Gamma$ is the transmission probability WKBJ tunneling probability $e^{-2\Delta \gamma}$. Thus an examination of $P_{SH}$ is also relevant to $P_R$. The tunneling power (Rabinowitz, 1999 a, b, c) radiated from a LBH is:

$$P_R \approx \left[\frac{\hbar c^3}{4\pi GM}\right]\frac{\langle\Gamma\rangle c^3}{4GM} = \left[\frac{\hbar c^6 \langle\Gamma\rangle}{16\pi G^2}\right]\frac{1}{M^2} \sim \frac{\langle\Gamma\rangle}{M^2}\left[3.42 \times 10^{35} W\right], \tag{34}$$

where M in kg is the mass of the LBH. Since $P_R$ is due to a tunneling process and is not an information voiding Planckian black body radiation distribution, it can carry information related to the formation of a BH, and avoid the information paradox associated with Hawking radiation.

Two LBH may get quite close for maximum tunneling radiation. In this limit, there is a similarity between the tunneling model, and what is expected from the Hawking model (1974,1975) in that the tidal forces of two LBH add together to give more radiation at their interface in Hawking's model, also producing a repulsive force.

## 8. WHY HAWKING RADIATION MAY BE UNDETECTABLE

In Hawking's model of black hole radiation, quantum field theory is superimposed on curved spacetime, and gravity is described classically according to Einstein's general relativity. It is a semiclassical approach in which only the matter fields are quantized, and black hole evaporation is driven by quantum fluctuations of these fields. After over a quarter of a century Hawking radiation has neither been detected experimentally, nor has intense theoretical effort succeeded in predicting the time history of black hole decay. This is the case with the present relatively elemental semi-classical approach, and a theory of quantum gravity is still not in sight.

### 8.1 Belinski: Hawking Radiation Does Not Exist

Belinski (1995), a noted authority in the field of general relativity, unequivocally concludes "the effect [Hawking radiation] does not exist." He comes to the same conclusion regarding Unruh (1976)-Davies (1975) radiation. He argues against Hawking radiation due to the infinite frequency of wave modes at the black hole horizon, and that the effect is merely an artifact resulting from an inadequate treatment of singularities. It is noteworthy that Unruh (1976) maintains that "an accelerated detector even in flat spacetime will detect particles in the vacuum. ... a geodesic detector near the horizon will not see the Hawking flux of particles."

Belinski probes deeper than this, since the fact that a derivation is invalid does not disprove the existence of an effect. He goes on to set up the problem with what in his terms are proper finite wave modes. He then concludes that no particle creation can occur.   As explained in section 7, the Rabinowitz radiation tunneling model (1999 a, b, c) involves no infinite frequency wave modes, and does not invoke the creation of particle-antiparticle pairs to produce black hole radiation.

Belinski determines that one reason the Hawking and Unruh effects are mathematical artifacts is that they violate a principle of quantum theory "which does not permit the physical particle wave functions to have singularities at those space-time points at which the external field is regular and where there are no sources."  This happens at the horizon in their derivations despite the fact that all horizon points are regular and free of any sources.  There have been attempts to resolve this problem with a high frequency cutoff.  However, this is antithetical to relativity theory, and is more questionable than the problem being rectified.

Many reputable scientists questioned the validity of the Hawking model (1974, 1975) not long after its introduction.   Belinski is not the only one to question the existence of Hawking radiation in recent times.  But his is a most cogent and most recent challenge.  Some of the other challenges have been both less manifest and less direct.  De Sabbata and Sivaram (1992)  suggest that "Thus one may observe the decay [Hawking radiation] only if one makes an infinite succession of measurements.  So in a sense one may never be able to observe the Hawking effect." Balbinot (1986) concluded that highly charged black holes do not radiate.  He concludes that "For an extreme Reissner-Nordstrom black hole ... there is no Hawking evaporation."  As a black hole becomes more and more charged, the Hawking radiation decreases until there is none. The maximum charge it can hold is when the electrical potential equals twice the gravitational potential, $z^2 e^2 / 4\pi\varepsilon r = 2GM^2 / r \Rightarrow z = (M/e)[8\pi\varepsilon G]^{1/2} = 7.6 \times 10^8 \, M_{kg}$. So a $10^{-3}$ kg (1gm) LBH could hold up to a maximum of $7.6 \times 10^5$ net electron or proton charges.   The ionization mechanism of section 14.1 is one way in which very small LBH could get charged in going through the atmosphere and lightning clouds to approach becoming extreme Reissner-Nordstrom black holes with significantly reduced

Hawking radiation, but with little effect on tunneling radiation. This may also have happened in the earlier more dense universe.

Belinski's first objection is akin to one that is called the "backreaction on the metric" raised by many researchers (Wald, 1977), but which from Belinski's perspective was not really laid to rest. Unruh and Wald (1984) raise and claim to resolve "Several paradoxical aspects of this process related to causality and energy conservation... ." As testimony that the backreaction problem is still in need of better understanding is the recent consideration of related anomalies (Balbinot, Fabbri, and Shapiro, 1999). They say, "Notwithstanding decades of intensive studies, the evolution and fate of an evaporating black hole (EBH) are still unknown." Notwithstanding their own assiduous effort, they conclude that attempts so far are not correct.

In the exposition of his second reason, Belinski shows that if done properly there is no real particle-antiparticle creation because of "The inability of the particle to cross the barrier between the two Dirac seas ...."
Infinities are manipulated in ways that are not justifiable. Since there is no experimental verification of Hawking or Unruh radiation, it can not be said that they are experimentally justified.

The problem of the backreaction on the metric can be understood in simple terms without having to consider infinities at the black hole horizon. The average emitted energy from a LBH is

$$\langle mc^2 \rangle \approx kT \approx k\left[\frac{\hbar c^3}{4\pi kG}\right]\frac{1}{M} = k[2.46 \times 10^{23}]\frac{1}{M} \quad (35)$$

with M in kg, and where the temperature T is given by Hawking (1974). If we let m = M, eq. (35) implies that $m \approx \left[\frac{\hbar c}{G}\right]^{1/2} = M_{Pl}$ = 2.18 x 10$^{-8}$ kg

10$^{-5}$ gm. Though Hawking does impose the limitation that a LBH has $M \gg M_{Pl}$, there seems to be no limitation that emitted particles have a mass considerably less than the LBH mass M. For example, even though the average emitted mass from a $10^{-4}$ gm LBH is only $4 \times 10^{-7}$ gm, the Hawking model allows radiation of particles with greater mass with a non-negligible backreaction on the metric. One group thinks this is not a problem. Belinski lucidly articulates the view of those that maintain that backreaction is a problem.

## 8.2 Effects of Compacted Dimensions

Another approach assumes the correctness of the Hawking model, but analyzes the effects of additional compact dimensions on the attenuation of this radiation. Argyres et al (1998) conclude that the properties of LBH are greatly altered and LBH radiation is considerably attenuated from that of Hawking's prediction. Their LBH are trapped by branes so essentially only gravitons can get through the brane, which may be thought of as an abbreviation for vibrating membrane. For them, not only is the radiation rate as much as a factor of $10^{38}$ lower, but it also differs in being almost entirely gravitons.

## 9. HIERARCHY AND COMPACTED SPACE

A framework has been proposed for unifying the weak gravitational force with the strong force by postulating the existence of 2 or more compact dimensions in addition to the standard 3 spatial dimensions that we commonly experience. In this view, gravity is strong on a scale with the higher-dimensional compacted space, and only manifests itself as being weak on a larger 3-dimensional scale.

Although modern hierarchy theory is independent of string theory, it borrows from and has much in common with string theory. It does not require the (9 spatial + 1 time) dimensions of string theory. It utilizes the same concepts of restricting other forces that reside inside the compacted dimensions to remain therein, while allowing the gravitational force to manifest itself from the compressed space into 3-space. A testable prediction of one version of this theory is that if there are two and only two additional dimensions there should be a deviation from the $1/r^2$ Newtonian force at sub-millimeter dimensions (Arkani-Hamed et al, 1998). As shown by eq. (14), in a 5-dimensional space, one may expect a $1/r^4$ dependence of the gravitational force.

The degree of arbitrariness in this hierarchy theory can be illustrated by its prediction of the size of the extra compacted dimensions

$$r_c \sim 10^{\frac{30}{d}-17} \text{ cm,} \qquad (36)$$

where d = n - 3. For d = 1 (4-space), eq. (9.1) predicts $r_c \sim 10^{13}$ cm $\sim 10^8$ miles. The distance of the earth to the sun is $9.3 \times 10^7$ miles. So there cannot be only one extra dimension, since the Newtonian gravitational force is well established at this scale. For d = 2 (5-space), both extra dimensions would have $r_c \sim 10^{-2}$ cm. For d = 3 (6-space), the three extra dimensions would all be at the atomic dimension $r_c \sim 10^{-7}$ cm. The 6 extra dimensions of string theory would all have $r_c \sim 10^{-12}$ cm, so the impact on gravity would be at the nuclear scale.

If there are only 2 compacted dimensions, so that $r_c \sim 10^{-2}$ cm, it may be possible for gravitational atoms of radius $r < \sim r_c$ to be protected from disintegration inside the compacted dimensions despite their low binding energy, which can be greater in higher dimensional space.

## 10. DARK MATTER

We know what only 5% of the accessible universe is made of. One piece of evidence that there must be 95% dark matter or missing mass comes from spiral galaxies. There must be some unseen form of matter whose gravitational attraction is great enough to hold the galaxies together as they rotate, as discovered by Vera Rubin (1983). The missing mass gives the stars ~ constant linear velocities independent of radial distance r, rather than the expected Keplerian velocities $\propto 1/\sqrt{r}$. Their rate of rotation is so great that they would fly apart if they contained only the stars and gas we can directly perceive.

Since LBH are extremely massive for their miniscule size, they may well explain the missing mass or so-called dark matter of which the universe is composed, as determined independently by galaxy rotation, galactic clusters, and the average density of the universe to account for its apparently Euclidean structure. LBH qualify as cold dark matter since their velocities $<< c$, as needed for early gravitational clustering. Since they can be small compared with the wavelength of visible light, they will not scatter or occlude light from the distant stars. For example, black holes of between $10^{-7}$ kg and $10^{19}$ kg have radii between $10^{-34}$ m and $10^{-8}$ m, well below visible wavelengths of $4 \times 10^{-7}$ m to $8 \times 10^{-7}$ m (4000 to 8000Å). To account for the missing dark matter there would need to be between $10^{61}$ and $10^{35}$ such black holes for a universe mass of $\sim 10^{53}$ kg. For our universe of radius $15 \times 10^9$ light-year, this would require an average density of between $10^{30}$ and $10^4$ black holes per cubic light-year ($10^{48}$ m$^3$) and more than this near the over 400 billion galaxies in the universe, due to gravitational attraction.

This is far greater than permitted for Hawking's extremely radiative LBH, which at most can make up $10^{-6}$ of the matter in the universe, or there would presently be too much radiation and they would have interfered with the nucleosynthesis of protons and neutrons into deuterium, helium, etc. in the early universe. In my model, LBH are much more quiescent than Hawking's, are much less likely to interfere with nucleosynthesis than his, and thus can account for the dark matter, i.e. up to 95% of the mass of the universe.

String theory suggests that dark matter is matter in parallel universes in other dimensions whose light cannot penetrate to reach us, but whose gravity can. This seems to be far more speculative than LBH which do not radiate according to the Hawking model.

## 11. BLACK HOLES AND ENTROPY OF THE UNIVERSE

In 3-dimensional space Bekenstein (1972 - 1974) found that the entropy of a black hole is

$$S_{bh} = kAc^3/4G\hbar = \frac{kc^3}{4G\hbar}R_H^2 = \frac{kc^3}{4G\hbar}\left[\frac{2GM}{c^2}\right]^2 = k\frac{M^2}{\left(\frac{c\hbar}{G}\right)}$$

$$= k\left[\frac{M}{M_{Pl}}\right]^2 = k\ln N \qquad (37)$$

where A is its surface area/$4\pi$ (neglecting the entropy due to wrinkling of the surface by other bodies), M is the mass of the black hole, $M_{Pl}$ = 2.18 x $10^{-8}$ kg is the Planck mass, k = 1.38 x $10^{-23}$ J/K, and k ln N is the standard Boltzmann statistical mechanical entropy of a system containing N distinct states.

It follows from his formulation that the entropy of black holes is tremendously greater than the entropy of ordinary bodies of the same mass. For example, our sun of mass 2 x $10^{30}$ kg (~ $10^{57}$ nucleons) and radius ~ $10^9$ m (~$10^6$ miles) has entropy $S \approx 10^{35}$ J/K, whereas a black hole of the same mass has entropy $S_{bh} \approx 10^{53}$ J/K, $10^{18}$ times higher with a radius of only ~ $10^3$ m (~1 mile). If the universe were 95% full of such black holes, there would be $10^{23}$ of them with a total entropy of $10^{76}$ J/K. This represents an excess entropy of $10^{41}$ times that of our universe if it were filled with stars like our sun. Thus there is a colossally higher probability that the big bang produced black holes dominantly over ordinary matter. This is a possible solution to the conundrum of why the early universe appears to have so little entropy. It appears likely that a large percentage of the mass of at least the primordial universe was composed of little black holes according to my model. This is particularly so, since interference with nucleosynthesis would no longer be an issue. One may well expect LBH to be a major constituent of the remnants of the big bang, but can't be according to Hawking.

The precise entropy increase over that presently inferred depends on the distribution of LBH masses and that of ordinary matter. The LBH entropy is sensitive to the mass distribution as it depends on $M^2$ per LBH. For example if we consider that 95% of the universe was initially composed of $10^{20}$ kg LBH with radius of only $10^{-5}$ cm, there would be ~ $10^{30}$ such LBH, each with entropy of $10^{33}$ J/K with a total entropy of $10^{66}$ J/K. This is still impressively high.

Thought experiments to test Bekenstein's entropy law were conducted. In this context Unruh and Wald (1982) considered ways to " 'mine' energy from a black hole. ... [and] Analogous effects for accelerating boxes in flat spacetime."

Generalizing Bekenstein's equation by using $R_{Hn}$ as given by eq. (29), the entropy of a black hole in n-space is

$$S_{bhn} \propto \frac{kc^3}{4G_n \hbar}[R_{Hn}]^2 = \frac{kc^3}{4G_n \hbar}\left[\frac{4\pi G_n M \Gamma(\frac{n}{2})}{(n-2)\pi^{n/2}c^2}\right]^{\frac{2}{n-2}}$$
$$\propto k\left[\frac{\Gamma(\frac{n}{2})}{(n-2)\pi^{n/2}}\right]^{\frac{2}{n-2}}\left[\frac{M}{M_{Pl}}\right]^{\frac{2}{n-2}} \quad . \quad (38)$$

It is noteworthy that the contribution to the entropy of the universe increases for smaller mass black holes as the dimensionality of n-space gets higher than 4.

If our universe were a black hole then its entropy would be

$$S_{bhn} \propto k\left[\frac{\Gamma(\frac{n}{2})}{(n-2)\pi^{n/2}}\right]^{\frac{2}{n-2}}\left[\frac{M_U}{M_{Pl}}\right]^{\frac{2}{n-2}} \quad . \quad (39)$$
$$\xrightarrow{n=3} k\left[\frac{M_U}{M_{Pl}}\right]^2 = k\left[\frac{10^{53}\text{kg}}{10^{-8}\text{kg}}\right]^2 = 10^{99}\text{J}/\text{K}$$

If $G_n$ were known, it would be possible to determine n, the dimensionality that would maximize the entropy of the universe treated as if it were a black hole.
The conundrum of why the early universe appears to have so little entropy may have a solution in that LBH would give it a large entropy.

Black hole "no-hair" theorems state that black holes can be completely characterized by a few variables such as mass, angular momentum, electric charge, and magnetic charge (monopoles).  Perhaps the minute wrinkling of a black hole's surface on the scale of the Planck area by gravitational perturbations due to external matter of the rest of the universe can contribute significantly to a black hole's entropy.  If it were not for "no-hair" theorems, it would appear from the above examples that for completeness if the universe were a black hole, one should also take into consideration the entropy inside it as well as the entropy associated with its horizon area; or conclude that the universe is not a black hole, i.e. that its space is Euclidean.  In 3-dimensions for a total mass M composed of N masses, the internal entropy $S_{3\,int} = Nk\left[\dfrac{M/N}{M_{Pl}}\right]^2 = \dfrac{k}{N}\left[\dfrac{M}{M_{Pl}}\right]^2$ decreases substantially as N gets large.  All the above seems to imply that the universe cannot be closed in the sense of being a black hole.

## 12. BLACK-BODY RADIATION IN n-SPACE

In 1879, Stefan empirically discovered the $T^4$ black-body radiation law, based on Tyndall's measurements.  Shortly after, Boltzmann derived the $T^4$ dependence by a purely thermodynamic argument.  The Stefan constant (now called the Stefan-Boltzmann constant) remained an experimentally determined proportionality constant, until Planck derived it theoretically from his black-body radiation law.

Let us generalize Boltzmann's derivation.  In n-space with n degrees of freedom, the radiation pressure $P_n = \frac{1}{n}u_n$, where $u_n$ is the energy density.  The internal energy $U_n = u_n V_n$, where $V_n$ is the n-volume.  The thermodynamic relation for internal energy is

$$\frac{\partial}{\partial V_n}(U_n)_T = T\left(\frac{\partial P_n}{\partial T}\right)_V - P_n \Rightarrow \frac{\partial}{\partial V_n}(u_n V_n) = T\frac{\partial}{\partial T}\left(\frac{u_n}{n}\right) - \frac{u_n}{n}. \quad (40)$$

Equation (40) leads to

$$\frac{du_n}{u_n} = (n+1)\frac{dT}{T} \Rightarrow u_n \propto T^{n+1}. \tag{41}$$

Thus the n-dimensional equivalent of the Stefan-Boltzmann black-body radiation law from eq. (41) is

$$P_{BBn} \propto cu_n \propto T^{n+1}. \tag{42}$$

## 13. LITTLE BLACK HOLES AND BALL LIGHTNING

Prior to the awareness that LBH radiate appreciably, their presence on earth was considered highly unlikely, as LBH would devour the earth ~ million years. But with radiation evaporation of LBH, their lifetime in the earth's vicinity < ~ year would be much less than the time it would take to ingest the earth. LBH would be unlikely on earth with Hawking radiation, because this radiation is devastating in all directions. The view of radiation from LBH presented by the Rabinowitz model (1999 a, b) obviates both of the above problems since this radiation is beamed and considerably less than Hawking's (1974, 1975). In the Rabinowitz model, when LBH get so small that there would be appreciable exhaust radiation, the radially outward radiation reaction force propels them away from the earth.

Ball lightning is widely accepted, but still unexplained. A testable LBH model for BL is presented which explains most of the known features of BL. In this model, LBH produce visible light in interacting with the atmosphere. The BL core energy source is gravitationally stored energy which is emitted as beamed radiation by means of gravitational field emission.

Most of the results in the following sections are derived independently of the model of black hole radiation. Near the LBH, exhaust radiation can augment ionization and excitation, but this complication will not be introduced at this time. Although a number of mechanisms are at work, polarization and ionization by the LBH gravitational and electrostatic tidal force is the major direct LBH interaction analyzed in this paper. LBH with mass ~ $10^{-3}$ kg and radius ~ $10^{-30}$ m are found to be the most likely candidates to manifest themselves as ball lightning (BL).

## 14. LBH GRAVITATIONAL AND ELECTROSTATIC TIDAL FORCE 14.1 Gravitationally Enhanced Ionization Cross-Section

The intense attractive converging gravitational and/or electrostatic field of a charged LBH causes more atmospheric molecules to be polarized and ionized than given by only kinetic considerations. Let us first examine the gravitational case. The gravitational potential energy of a particle of mass m in the field of a LBH is

$$V = -\frac{GMm}{r} - p\left(\frac{GM}{r^2}\right) - \frac{\alpha_p}{2}\left(\frac{GM}{r^2}\right)^2, \qquad (43)$$

where p is the permanent dipole moment, which will usually be negligible for atoms but not for molecules, and $\alpha_p$ is the gravitational polarizability. We will be dealing primarily with atoms of the disassociated molecule since the binding energy of the molecules << the ionization potential, and they will be torn apart well before getting in close enough for ionization. When the atomic collision frequency is low compared with the ionization rate due to tidal interaction with the gravitational field of the LBH, the ionization radius $r_i$ can be increased. This results in an enhanced ionization volume, i.e. an enhanced ionization cross-section $\sigma_E$.

To a first approximation, this problem will be treated as a simple central force problem in which angular momentum is conserved. Implications of (1) atomic scattering, (2) ionization and scattering by the LBH exhaust, and (3) tidal force interactions will be neglected for now. These make orbital motion non-reentrant about the LBH as indicated by the Runge vector (or Runge-Lenz vector, quantum mechanically). Scattering is negligible as an LBH enters the low density atmosphere from outer space and starts to produce ions around it, and as we shall see even at high density, when the mean free path $\lambda$ > the enhanced interaction radius as calculated in this section. The interaction analysis here applies to both the sphere of ionization and to the sphere of polarization. So the symbol $r_{ip}$ will represent either the ionization radius or the larger polarization radius depending on which case is to be considered.

We can make the problem one-dimensional by introducing an effective potential energy

$$V_{eff} = V(r) + \frac{L^2}{2mr^2}, \qquad (44)$$

where L is the conserved angular momentum of an atom about the LBH.

$$L = mvr_E = m(\Im kT/m)^{1/2} r_E = (\Im mkT)^{1/2} r_E, \qquad (45)$$

where T is the temperature of the gas, $\Im \geq 3$ is the number of degrees of freedom of the particle, and $r_E$ is the enhanced ionization radius, i.e. the new larger radius of furthest approach for ionization of an atom.

The radial velocity $v_r = [2(E - V_{eff})/m]^{1/2} = 0$ at the closest approach to $r_i$, for a particle that just grazes the original ionization sphere. Hence at $r = r_{ip}$, $V_{eff} = E = \frac{\Im}{2} kT$. Combining this with equations (43) and (44) yields

$$V(r_{ip}) = \tfrac{\Im}{2} kT \left[1 - \left(r_E/r_{ip}\right)^2\right]. \qquad (46)$$

Therefore equation (46) gives us the new enhanced ionization radius,

$$r_E = r_{ip}\left[1 - \frac{1}{\tfrac{\Im}{2}kT} V(r_i)\right]^{1/2}. \qquad (47)$$

The ionization-polarization radius increases since V is negative.

We next need to determine the gravitational polarizability $\alpha_p$. The gravitational tidal force $F_T$ polarizes an atom,

$$F_T = \frac{(ze)^2 \partial}{4\pi\varepsilon a^3}, \qquad (48)$$

where a is the unperturbed atomic radius, $\partial$ is the displacement relative to the electron cloud of the nucleus of mass $m_N \approx m$ the atomic mass, and $\varepsilon$ is the permittivity of free space. The term on the right is the electrical harmonic restoring force with spring constant $K = (ze)^2/4\pi\varepsilon a^3$. The displacement produces both electric and gravitational dipole moments, of which the latter is

$$m_N \partial \approx m\partial = \alpha_p (F_T/m). \tag{49}$$

Combining equations (48) and (49) yields a general result independent of the form of $F_T$.

$$\alpha_p = \left(\frac{4\pi\varepsilon m^2}{(ze)^2}\right) a^3. \tag{50}$$

Substituting equation (50) into (43),

$$\begin{aligned} V &= -\left(\frac{GMm}{r}\right) - p\left(\frac{GM}{r^2}\right) - \frac{1}{2}\left(\frac{4\pi\varepsilon m^2 a^3}{(ze)^2}\right)\left(\frac{GM}{r^2}\right)^2 \\ &= -\left(\frac{GMm}{r}\right) - p\left(\frac{GM}{r^2}\right) - \left[2\pi\varepsilon\left(\frac{GMm}{ze}\right)^2\right]\frac{a^3}{r^4} \end{aligned} \tag{51}$$

Thus from equation (47), the gravitationally enhanced ionization cross-section of the LBH is

$$\sigma_{gE} = \pi r_{ip}^2 \left\{1 + \frac{1}{\frac{3}{2}kT}\left[\frac{GMm}{r_{ip}} + p\left(\frac{GM}{r_{ip}^2}\right) + 2\pi\varepsilon\left(\frac{GMm}{ze}\right)^2 \frac{a^3}{r_{ip}^4}\right]\right\}. \tag{52}$$

Equation (52) is applicable if the particle mean free path $\lambda < r_E$. For convenience, this will be called the low density case. Whether the low or high density case is relevant is a function of both the density of the gas and the mass M of the LBH, since for small $r_E$, the mean free path $> r_E$ even above atmospheric pressure.

### 14.2 Electrostatically Enhanced Ionization Cross-Section

A similar analysis can be done for the electrostatic case. The resulting electrostatically enhanced ionization cross-section of a charged LBH is

$$\sigma_{eE} = \pi r_{ip}^2 \left\{1 + \frac{1}{\frac{3}{2}kT}\left[\frac{Qq}{4\pi\varepsilon r_{ip}} + p\left(\frac{Q}{4\pi\varepsilon r_{ip}^2}\right) + 2\pi\varepsilon\left(\frac{Q}{4\pi\varepsilon}\right)^2 \frac{a^3}{r_{ip}^4}\right]\right\}, \tag{53}$$

where Q is the electric charge of the LBH, q is the net charge of the atom or molecule, and the electric polarizability $\alpha_{pe} = 4\pi\varepsilon a^3$. Though only single ionization will be considered, higher degrees of ionization are possible.

## 15. BALL LIGHTNING RADIATION

### 15.1 Ionization Rate

The ionization rate due to a LBH moving through the atmosphere is

$$\frac{dn_i}{dt} \sim \sigma\bar{v}n^2 + \sigma v_{BL}n^2 - \sigma_{re}\bar{v}n_i^2 - \sigma_{diff}\bar{v}n_i^2, \tag{54}$$

where n is the number density of atoms, $\sigma$ is the ionization cross-section (enhanced or unenhanced depending on relative mean free path), $\sigma_{re}$ is the recombination cross-section, $\sigma_{diff}$ is the cross-section for diffusion out of the ionization sphere, $\bar{v}$ is the mean thermal velocity, $v_{BL}$ is the BL velocity, and $n_i$ is the number density of ions. The solution of equation (54) is

$$n_i \sim \left[\frac{A(Be^{Ct}-1)}{(Be^{Ct}+1)}\right] \underset{t\to\infty}{\approx} A = \left(\frac{\sigma(\bar{v}+v_{BL})n^2}{\sigma_t\bar{v}}\right)^{1/2} \tag{55}$$

where $B = (A + n_{io})/(A - n_{io})$, $n_{io}$ is the initial number density of ions, $\sigma_t = \sigma_{re} + \sigma_{diff}$, and $C = 2n[\sigma\sigma_t(\bar{v}+v_{BL})\bar{v}]^{1/2}$.

### 15.2 Recombination Radiation

As the LBH moves through the atmosphere, its gravitational and/or electrostatic tidal force excites and ionizes air atoms around it and carries the generated plasma along by electrostatic and/or gravitational attraction. At early times, the ionization time is short relative to the recombination time, and to the time for diffusion out of the ionization sphere. Entry into a LBH is difficult since the particle's deBroglie wavelength needs to be $< \sim R_H$ and because of conservation of angular momentum. In the presence of the LBH gravitational field and gradient, both the recombination and the diffusion times are longer than in free space, and $\sigma_{re}$ is reduced.

The equilibrium solution is obtained from Equation (55) as t gets large. In this limit the recombination rate is

$$R_{re} = [\sigma_{re}\overline{v}n_i^2] \sim \sigma(\overline{v}+v_{BL})n^2 \left\{ \left( \frac{\sigma\sigma_{re}}{(\sigma_{re}+\sigma_{diff})^2} \right) \left( \frac{\overline{v}+v_{BL}}{\overline{v}} \right) \right\}. \quad (56)$$

The radiation is hardly perceptible at first. In steady state, the electron-ion recombination radiated power/volume is

$$P_{re} = [R_{re}]V_i \sim \left[ \sigma(\overline{v}+v_{BL})n^2 \left\{ \left( \frac{\sigma\sigma_{re}}{(\sigma_{re}+\sigma_{diff})^2} \right) \left( \frac{\overline{v}+v_{BL}}{\overline{v}} \right) \right\} \right] V_i. \quad (57)$$

$V_i$ is the ionization potential (15.5 eV for nitrogen), and n is the number density of air atoms. For the LBH mass range of interest ~1 gram, equation (57) yields > ~ Watts of radiated power in agreement with observation. Photons with 15.5 eV energy have a frequency higher than visible photons of a few eV with wavelengths between 4000 Å and 8000 Å. However, energy degradation and other radiation mechanisms can result in visible light. There is comparable thermal and de-excitation radiation.

A less detailed impulse transfer approach yields a power transfer of

$$P = \frac{4\pi G^2 M^2 \rho}{v_{bh}} \ell n \left( \frac{b_{max}}{b_{min}} \right) \sim 10 \text{ W}$$ to the atmosphere by a LBH with M ~ $10^{12}$ kg, $R_H$ ~ $10^{-15}$ m, $\rho_{atm}$ = 1.3 kg/m³ is the atmospheric mass density, and the weak logarithmic dependence of the ratio of the maximum to minimum impact parameters $\ell n(b_{max}/b_{min}) \sim 30$.

## 16. QUASI-ORBITAL NEAR-CAPTURE ELECTRON RADIATION

A LBH with charge Q can form a superheavy atom/ion. Appreciable radiation will ensue when a quasi-orbiting particle of net charge $q_{net}$ falls in toward the LBH and is nearly captured. Energy E is radiated per particle at frequency $f_o$, where the initial orbital velocity $v_o$ $2\pi f_o r_o$. The emission of radiation near $f_o$ can account for the BL colors of low optical power:

$$E \sim \frac{Qq_{net}}{6\pi^2\varepsilon_o c}\left(\frac{v_o}{c}\right)^2 \int_{f_{min}}^{f_{max}} \left[\frac{f^2(f_o^2 + f^2)}{(f^2 - f_o^2)^2}\right] df$$

$$= \frac{Qq_{net}}{6\pi^2\varepsilon_o c}\left(\frac{v_o}{c}\right)^2 \left[f_o \ell n\left(\frac{f-f_o}{f+f_o}\right) + \frac{f(f^2 - 2f_o^2)}{f^2 - f_o^2}\right]_{f_{min}}^{f_{max}}$$

(58)

## 17. BEAMED LBH RADIATION CAN PRODUCE LEVITATION

The downwardly directed radiation (due to the earth below) from a $3 \times 10^{-4}$ kg ( 1/3 gm) LBH will act like a rocket exhaust permitting the LBH to levitate or fall slowly. We can estimate the upward force on the LBH from

$$M\frac{dv}{dt} = -c\frac{dM}{dt} - Mg \qquad (59)$$

where the exhaust leaves the LBH at near the speed of light, $c = 3 \times 10^8$ m/sec, the acceleration of gravity $g = 9.8$ m/sec$^2$ near the earth's surface, and $dM/dt = -P_R/c^2$. For levitation $P_R \sim 10^6$ W. In one model, emission is mainly by the six kinds of neutrinos (Thorne et al., 1986) and in another almost entirely by gravitons (Argyres et al., 1998). The emitted power $P_R$, necessary to produce levitation, as well as the necessary masses and separations of the LBH and host body needed to produce this exhaust power are independent of the nature of the emitted particles. At a distance of many earth radii, the radiation is narrowly beamed toward the earth's center. As a LBH gets close to the earth the radiation beam diverges to ~ the earth's diameter, giving it a low power density

## 18. INCIDENCE RATE OF BALL LIGHTNING

The continuity equation for mass flow of LBH when there is a creation rate $S_c$ and a decay rate $S_d$ of mass per unit volume per unit time t is

$$\nabla \bullet (\rho\vec{v}) + \partial\rho/\partial t = S_c - S_d, \qquad (60)$$

where $\rho$ is the LBH mass density at a given point in the universe, $\vec{v}$ is the LBH velocity, and $\rho\vec{v}$ is the LBH flux density. In steady state, $\partial\rho/\partial t = 0$. Integrating eq. (60):

$$\int (\rho \vec{v}) \bullet d\vec{A} = \int (S_c - S_d) dV_t \Rightarrow$$
$$-\rho_{LBH} v_{LBH} A_{far} + \rho_{BL} v_{BL} A_E = (S_c - S_d) V_t' \tag{61}$$

where $\rho_{LBH}$ is the mass density of LBH at a distance far from the earth, typical of the average mass density of LBH throughout the universe. $A_{far}$ is the cross-sectional area of a curvilinear flux tube of LBH far from the earth, $A_E$ is the cross-sectional area of the tube where it ends at the earth, and $V_t$ is the volume of the curvilinear flux tube (cylinder). Since the LBH were created during the big bang, at a large distance from the earth they should be in the cosmic rest frame. The velocity of our local group of galaxies with respect to the microwave background (cosmic rest frame), $v_{LBH} \sim 6.2 \times 10^5$ m/sec (Turner and Tyson, 1999), is a reasonable velocity for LBH with respect to the earth.

Because $v_{LBH}$ is high and LBH radiate little until they are near other masses, $S_c$ can be neglected with negligible decay of large black holes into LBH in the volume $V_t$. Similarly, $S_d$ may be expected to be small until LBH are in the vicinity of the earth where most of their evaporation, before they are repelled away, is in a volume of the atmosphere $\sim A_E h$, where $A_E$ is the cross-sectional area of the earth, and h is a characteristic height above the earth. At this point it is helpful to convert to number density $\rho_L$ and $\rho_B$, of LBH and ball lightning respectively. The number density decay rate is $\rho_B A_E h/\tau$, where $\tau < \sim$ year is the dwell-time of LBH near the earth. Thus eq. (61) yields

$$\rho_B = \rho_L \left[ \frac{v_{LBH}}{v_{BL} + (h/\tau)} \right] \frac{A_{far}}{A_E}, \tag{62}$$

which implies that the ball lightning flux is

$$\rho_B v_{BL} = \rho_L v_{LBH} \left[ \frac{v_{BL}}{v_{BL} + (h/\tau)} \right] \frac{A_{far}}{A_E} \approx \rho_L v_{LBH} \left( \frac{A_{far}}{A_E} \right), \tag{63}$$

where in most cases $h/\tau << v_{BL}$.

At large velocities, LBH that do not slow down appreciably due to their large mass or angle of approach, either do not produce sufficient ionization to be seen or do not spend sufficient time in the atmosphere to be observed. In the Rabinowitz model (1999a, b, c), those LBH that reach the earth's atmosphere and are small enough to have sufficient radiation reaction force to slow them down to the range of $10^{-2}$ to $10^2$ m/sec, with a typical value $v_{BL} \sim 1$ m/sec, manifest themselves as BL. So eq. (63) implies that the ball lightning current in the atmosphere ≈ the LBH current far away. We can thus give a range for the BL flux density

$$\rho_L v_{LBH} < \rho_B v_{BL} < \rho_L v_{LBH}\left(\frac{A_{far}}{A_E}\right). \tag{64}$$

The distribution of LBH masses is not known. Assuming that LBH comprise all of the dark matter, i. e. 95 % of the mass of the universe (Rabinowitz, 1990 b) of which there is a percentage p of LBH of average mass $\overline{M}_{LBH} \sim 10^{-3}$ kg:

$$\rho_L \sim \frac{p(0.95 M_{univ} / \overline{M}_{LBH})}{V_{univ}}. \tag{65}$$

For $M_{univ} \sim 10^{53}$ kg, $V_{univ} \sim 10^{79}$ m$^3$ (radius of 15 x$10^9$ light-year = 1.4 x $10^{26}$ m), and p ~ 10 %, $\rho_L \sim 10^{-24}$ LBH/m$^3$. Thus from eqs. (64) and (65) my model predicts that the incidence rate of BL is roughly in the range
$10^{-12}$ km$^{-2}$ sec$^{-1}$ to >~ $10^{-8}$ km$^{-2}$ sec$^{-1}$ for $A_{far}/A_E >$ ~$10^4$. (Even if p = 100%, $10^{-11}$ km$^{-2}$ sec$^{-1}$ to $10^{-7}$ km$^{-2}$ sec$^{-1}$ is well below the signal level of existing detectors.) This rate is in accord with the estimates of Barry and Singer (1988) of 3 x $10^{-11}$ km$^{-2}$ sec$^{-1}$, and of Smirnov (1993) of 6.4 x $10^{-8}$ km$^{-2}$ sec$^{-1}$ to $10^{-6}$ km$^{-2}$ sec$^{-1}$. This is well below the incidence rate of lightning (Turman, 1977).

## 19. GENERAL DISCUSSION

### 19.1 Hierarchy Theory

Hierarchy theory is considerably older than string theory. The latter has its origins in an unsuccessful attempt to understand the strong force in the late 1960's. It appears that hierarchy started when Dirac (1937) proposed that the gravitational constant G decreases with the age of the universe so that $G \propto T_U^{-1}$. His suggestion was based upon a numerological observation that dimensionless ratios of some physical quantities yield very large numbers ~ $10^{40}$. For example, the ratio of the electric force to the gravitational force between two electrons is 4.2 x $10^{42}$, and is 1.2 x $10^{36}$ between two protons. It is interesting that Dirac (1938) took interest in dimensionless constants, since he took exception to Milne's "Dimensional Hypothesis" that constants with dimensions should not appear in cosmology, and to Milne's conjecture that $G \propto T_U$. In this 1938 paper, Dirac concluded, "We are thus left with the case of zero curvature, or flat t-space, as the only one consistent with our fundamental principle and with conservation of mass." This was a prescient conclusion, coming long before the very recent cosmic microwave background evidence that space-time appears to be flat on a large scale.

Dicke (1961) challenged Dirac with an anthropic argument, "… $T_U$ is not a random choice … but is limited by the criteria for the existence of physicists." Dicke further argued Mach's principle requires $G \propto T_U$. Dirac (1961) gave the following biocentric reply, "On this assumption, [Dicke's] habitable planets could exist only for a limited period of time. With my assumption, they could exist indefinitely in the future, and life would never end. There is no decisive argument for deciding between these assumptions. I prefer the one that allows the possibility of endless life." Dirac (1973) continued to maintain that G must have been larger in the early universe despite his acknowledgment, "Now Einstein's theory of gravitation requires that G shall be constant. … Thus Einstein's
theory of gravitation is irreconcilable with the Large Numbers hypothesis."

In more recent times, Gribb (1989) argues for an "effective change of the gravitational constant [at the time of the big bang] which leads to the possibility of the creation of particles with mass of the order of the observable universe … and now only particles with a microscopic mass can be created from the vacuum, so there can be no big bang now." In their (1974) solution to Dirac's Large Number Hypothesis, Georgi, Quinn, and Weinberg show that at very high energies (~$10^{16}$ GeV) the strong, weak, and electromagnetic forces come together in strength -- a very fundamental finding. They comment on gravity, "Perhaps gravitation has something to do with the superstrong spontaneous symmetry breaking." However present multidimensional hierarchy models would make their finding serendipitous rather than fundamental.

**19.2 String Theory and Compacted Dimensions**

Implications of higher dimensional space have been analyzed in my paper with neither approbation nor disapproval. The conclusions of hierarchy/string theory of a sub-millimeter compaction size do not appear to be compelling. The predictions regarding the size of the compacted dimensions can be modified down to the Planck length of $10^{-35}$ m, if experiment shows no deviation from standard Newtonian gravity at larger sizes. So far no deviation has been found down to 0.15 mm -- almost ruling out the 5-dimensional space predictions given in Sec. 9.

The extra dimensions are confined by branes. Until now the size of branes seemed to be so small that they would not contradict experimental findings since other forces have been probed to sub-nuclear sizes. The investigation for a deviation from Newtonian gravity is spurred by allowing branes to be ~ $10^{-1}$ mm in radius. If it is found at this scale, a deviation of the Coulomb force may also occur, which would seem to be in contradiction with established experiments at much smaller scales. There are also problems with conservation of energy.

Nevertheless string theory avoids long-standing singularity enigmas. Although the big bang is generally accepted, its initial singularity problem continues to be troubling. General relativity implies that at time zero, the universe was a point of zero volume with infinite density, temperature, etc. This singularity is circumvented by using string theory in which conventional point-like particles are replaced by one-dimensional strings having a very short length. Not only are the troublesome infinities disposed of, but string theory can endow the universe with a past before the big bang. Previously we were told that there was no "before." Now we may view the evolution leading to the big bang in much the same way as the explosion or collapse of stars.

### 19.3 Gravitational Strength

$G_n$ is model dependent and not necessarily a simple function of $r_c$. This would occur even if $G_n$ could properly (not arbitrarily) be made dimensionless, and is not a result of having units of

force(length)$^{n-1}$/(mass)$^2$.

Though the results presented here are for Euclidean space, the curved space of general relativity changes them only by dimensionless quantities which are expected to be small when $r > 10\, R_H$, as discussed in section 5.

### 19.4 Ball Lightning

Greatly decreased radiation relative to that in the Hawking model permits LBH to be prevalent throughout the universe. So it is reasonable to surmise that they are also present in the region of the earth, and may manifest themselves as BL. An interesting related question is how LBH manifest themselves in the region (accretion disk) around very large black holes where they may help initiate ionization, and provide a distinctive signature as they fall into a big black hole. In my radiation model, primeval LBH of small mass may still exist. However, even in the Hawking radiation model, small mass LBH may presently exist that started out as much more massive primeval black holes.

The following criteria are presented as a guide for assessing ball lightning/earth light models in general, and the little black hole model in particular. These are derived from several sources (Fryberger, 1994; Singer, 1971; Smirnov, 1993; and Uman, 1968). LBH meet most of the criteria for BL (Rabinowitz, 1999a, b): 1. Constant size, brightness, and shape for times < ~10 sec. 2. Untethered high mobility.

3. Generally don't rise.  4. Can enter open or closed structures.  5. Can exist within closed conducting metal structures such as airplanes and submarines.

6. Levitation.  7. Low power in the visible spectrum.  8. Rarity of sightings.  9. Relatively larger activity near volcanoes.  10. Abate quietly.  11. Extinguish explosively occasionally.  12. Related radioactivity.  13. Typical absence of deleterious effects. 14. Occasional high localized energy deposition.  15. Larger activity associated with thunderstorms. 16. Bounce.  17. Mass and Velocity Ranges.  18. Go Around Corners.

Ball lightning may extinguish by different mechanisms that relate either to the LBH or to the ionized atmosphere around the LBH. One gentle mode may simply be when the sphere surrounding the LBH becomes optically opaque, or the LBH enters an opaque medium. Another more violent mode may be when this sphere blows apart because the energy dissipation in the sphere can only go out at a limited rate and becomes too great to absorb into the LBH because of conservation of angular momentum; or if the LBH ingests too much charge. The result is likely an explosive release of the energy. Other modes are when the LBH is not an effective ionizer of the atmosphere; the LBH is repelled away from earth by the exhaust radiation; or it explodes catastrophically.

## 20. CONCLUSION

Both interesting and anti-intuitive results have been found regarding n-space and compacted dimensions regarding the gravitational force, quantized gravitational atoms and their radiation, entropy, black-body radiation, and LBH. The visible and total radiation from an alternate LBH tunneling radiation model was examined.

Gravitational and/or electrostatic tidal interaction of a LBH with air atoms can account for the reported visible radiation of BL. The incidence rate of LBH on earth has been shown to approximately equal the accepted assessment for the incidence rate of BL. The hypothesis that ball lightning is the manifestation of astrophysical LBH as they interact with the earth's atmosphere, deserves serious consideration.

**ACKNOWLEDGMENT**

I am grateful to Michael Peskin, Mark Davidson, Steve Crow, Arthur Cohn, and Elliott Bloom for helpful discussions.